\documentclass[journal,onecolumn]{IEEEtran}
\usepackage{cite}

\ifCLASSINFOpdf
\else
\fi
\usepackage{subfig}
\usepackage{mathtools}        
\usepackage{amssymb}
\usepackage{multirow}

\def\bmu{\mbox{\boldmath $\mu$}}

\begin{document}
%
\title{Unsupervised Pixel-wise Hyperspectral Anomaly Detection via Autoencoding Adversarial Networks}
%
%
%

\author{Sertac~Arisoy,
        Nasser~M. Nasrabadi,~\IEEEmembership{Fellow,~IEEE,}
        and~Koray~Kayabol,~\IEEEmembership{Senior Member~IEEE}
\thanks{This work is supported by the Scientific and Technological Research Council of Turkey (TUBITAK) under Project No. 118E295. The work of S. Arisoy was supported by the Council of Higher Education (YOK) of Turkey through the Research Abroad Scholarship (YUDAB) program.}
\thanks{S. Arisoy and Koray Kayabol are with the Electronics Engineering Department, Gebze Technical University, Gebze, Kocaeli, Turkey,
	e-mail: sarisoy@gtu.edu.tr, koray.kayabol@gtu.edu.tr.}
\thanks{N. M. Nasrabadi is with the Lane Department of Computer Science and Electrical Engineering, West Virginia University, Morgantown, West Virginia, USA, e-mail: nasser.nasrabadi@mail.wvu.edu.}
}

\maketitle

\begin{abstract}
We propose a completely unsupervised pixel-wise anomaly detection method for hyperspectral images. The proposed method consists of three steps called data preparation, reconstruction, and detection. In the data preparation step, we apply a background purification to train the deep network in an unsupervised manner. In the reconstruction step, we propose to use three different deep autoencoding adversarial network (AEAN) models including 1D-AEAN, 2D-AEAN, and 3D-AEAN which are developed for working on spectral, spatial, and joint spectral-spatial domains, respectively. The goal of the AEAN models is to generate synthesized hyperspectral images (HSIs) which are close to real ones. A reconstruction error map (REM) is calculated between the original and the synthesized image pixels. In the detection step, we propose to use a WRX-based detector in which the pixel weights are obtained according to REM. We compare our proposed method with the classical RX, WRX, support vector data description-based (SVDD), collaborative representation-based detector (CRD), adaptive weight deep belief network (AW-DBN) detector and deep autoencoder anomaly detection (DAEAD) method on real hyperspectral datasets. The experimental results show that the proposed approach outperforms other detectors in the benchmark. 

\end{abstract}


%
\IEEEpeerreviewmaketitle

\section{Introduction}
%
%
%
%
\IEEEPARstart{H}{yperspectral} anomaly detection (AD) is used in many applications such as surveillance, human-made object detection, and mine detection. Anomalous objects compose of small areas in the image that show different spectral characteristics compared to the neighboring background pixels. Due to the lack of a priori information about the anomaly pixels, in most of the anomaly detectors, the background model is estimated from the data and any pixel that does not fit the background model is assigned as an anomaly. In the literature, several anomaly detection approaches based on statistical, subspace, linear mixing, and deep neural network models are proposed. The Reed-Xiaoli (RX) AD \cite{RX90} is known as the benchmark of hyperspectral AD algorithms. In the RX algorithm, the background is statistically modeled with a multivariate Gaussian distribution. Besides the statistical approaches, there are also deterministic approaches for AD. A support vector data description (SVDD) AD is proposed in \cite{banerjee2006support} where the background is modeled with a minimum enclosing hypersphere. In \cite{li2014collaborative}, a collaborative representation-based detector (CRD) has been proposed, where the difference between a pixel and its collaborative representation is used for anomaly assignment.

Deep learning methods have also been utilized for hyperspectral AD. In \cite{bati2015hyperspectral}, an autoencoder model is proposed for representation learning and the spectral vector reconstruction via encoder and decoder, respectively. The reconstruction error between the original and the reconstructed spectral vectors is used as an anomaly score. Similarly, in \cite{taghipour2019unsupervised}, a deep autoencoder-based method called deep autoencoder anomaly detection (DAEAD) is proposed. In \cite{ma2017dbn}, a deep belief network (DBN) is trained to reconstruct the spectral vectors and the reconstruction error is used as an anomaly score. In \cite{ma2018unsupervised}, the contamination of anomaly pixels is reduced with an adaptive weight DBN (AW-DBN) detector. In \cite{li2017transferred}, a convolutional neural network (CNN)-based anomaly detector is proposed.

Among the deep learning approaches, the variants of recently introduced generative adversarial network (GAN) model in \cite{goodfellow2014generative} has been applied for anomaly/novelty detection \cite{eide2018applying, TJiang20, KJiang20, xie2019spectral, sabokrou2018adversarially, akcay2018ganomaly}. In \cite{eide2018applying}, Wasserstein GAN (WGAN) model is applied to hyperspectral AD. In \cite{TJiang20} and \cite{KJiang20}, the anomaly scores are calculated with the combination of the spectral and spatial information. In the spectral domain, GAN is used to estimate the background distribution. In the spatial domain, a morphological filter is used to extract anomaly scores. In \cite{xie2019spectral}, a feature representation of HSI is learned from the adversarial autoencoder (AAE) model and a morphological filter is applied for the background suppression.

In this paper, we propose a new approach for hyperspectral AD based on deep adversarial learning for background reconstruction and WRX \cite{guo2014weighted} for detection. The most similar works to our proposed approach are  \cite{sabokrou2018adversarially} and \cite{akcay2018ganomaly}. Our method differs from \cite{sabokrou2018adversarially} and \cite{akcay2018ganomaly} in several ways. The AD method in \cite{sabokrou2018adversarially} contains a denoising autoencoder and is applied for outlier and video anomaly detection problems. Instead of denoising autoencoder, we use an autoencoder in our proposed network model. In \cite{akcay2018ganomaly}, a model called GANomaly is used for semi-supervised anomaly detection from labeled image datasets. While three loss functions are used in \cite{akcay2018ganomaly} to train the network in a semi-supervised manner, we use two loss functions to train the network in an unsupervised manner. Unlike \cite{sabokrou2018adversarially} and \cite{akcay2018ganomaly}, we use an autoencoder under the adversarial learning framework for pixel-wise anomaly detection from hyperspectral images.

In summary, we propose a fully unsupervised pixel-wise anomaly detection method for hyperspectral images. For this purpose, we propose three different autoencoding adversarial network (AEAN) models for spectral (1D-AEAN), spatial (2D-AEAN), and joint spectral-spatial (3D-AEAN) domains. In the 1D-AEAN model, we apply only the spectral vectors to the input of the AEAN model to learn the background. Considering that, in practice, the background contains a number of different kinds of spectral vectors, learning them with a single 1D-AEAN model may be inefficient. Therefore, unlike the GAN models presented in \cite{TJiang20, KJiang20, xie2019spectral}, we utilize spatial information within the local area to generate synthesized images by 2D-AEAN and 3D-AEAN models. For this purpose, we train the 2D-AEAN model with extracted image blocks from each spectral channel. In the 3D-AEAN model, we combine the spectral and spatial information by training the 3D-AEAN model with small image cubes.
\begin{figure}[t]
	\centering{
		\includegraphics[width=9.3cm]{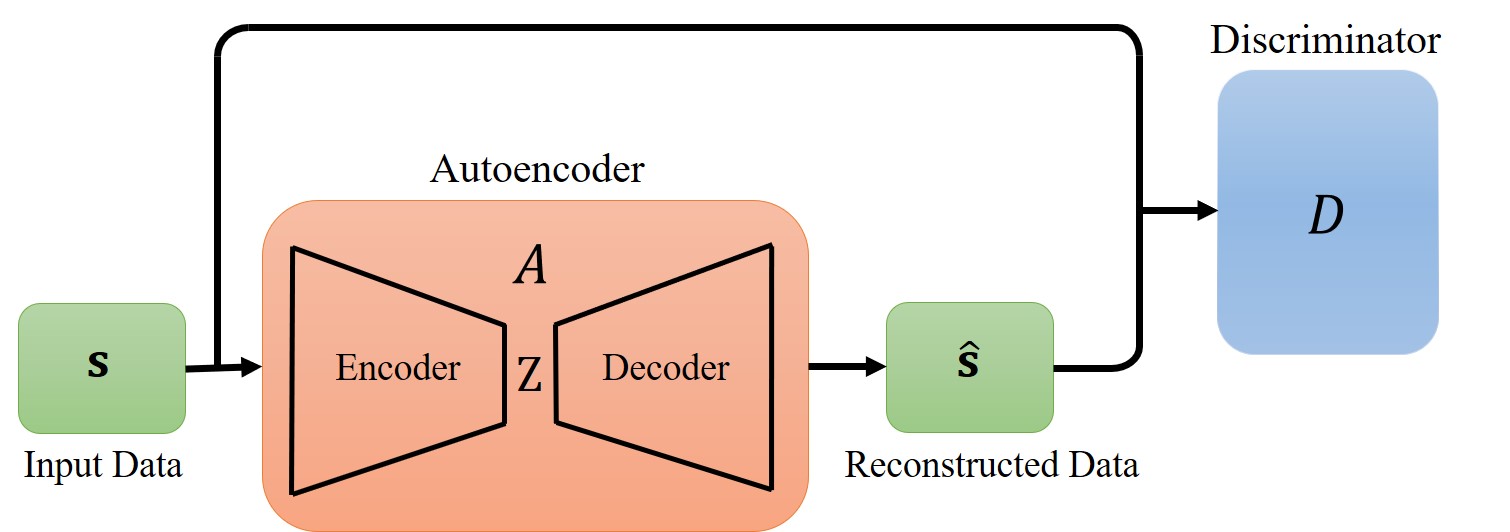}		
	}%
	\caption{The AEAN model, the input data can in be in the spectral, spatial or spectral-spatial domain.}
	\label{ganmodel}
\end{figure}
In the detection part, we propose to use a WRX-based detector \cite{guo2014weighted} for pixel-wise detection. In this model, if the spectral vector has similar characteristics with the background vector, its impact is increased, otherwise reduced. In \cite{ma2018unsupervised}, the impact of the anomaly pixels is reduced by an adaptive weight approach. The weights are calculated from the reconstruction error of the DBN. Unlike \cite{ma2018unsupervised}, in this paper, we propose to obtain the weights of the WRX detector using the reconstruction error of the AEAN. We also propose to apply background purification where likely background samples are extracted from the image to obtain a fully unsupervised method.

The organization of the paper is as follows: Section \ref{SecGAN} presents the proposed hyperspectral AD framework. Experimental results are reported in Section \ref{SecExp}. Section \ref{SecCon} summarizes the conclusion and future work.

\section{Proposed Anomaly Detection Framework}
\label{SecGAN}

In this section, we explain the proposed framework including data preparation, reconstruction, and detection steps.

\subsection{Data Preparation Step: Background Purification} \label{SecPure}

In this section, we remove the potential anomaly pixels from the image by using probability-based background purification method \cite{Zhao17} and prepare the training samples for the AEAN models. For this purpose, we calculate an anomaly score with Mahalanobis distance (MD). Let us denote HSI as $\mathbf{F} \in \mathbb{R}^{M_1\times M_2 \times L}$ where $L$ is the number of spectral bands and $M_1\times M_2$ is the size of the image at each spectral band. By denoting the spectral vector at $i$th pixel as $\mathbf{f}_i \in \mathbb{R}^{L}$, where $i = 1, \dots, N$ is the lexicographically ordered pixel index, the MD score evaluated at $i$th pixel is calculated as $\mathrm{MD(\mathbf{f}}_i)=(\mathbf{f}_i-\mathbf{\bmu})^{T}{\mathbf{\Sigma}}^{-1}(\mathbf{f}_i-\mathbf{\bmu})$, where $\bmu$ and $\mathbf{\Sigma}$ are the global sample mean and covariance matrix, respectively. We expect that the MD scores for the anomaly pixels are greater than the background pixels. In \cite{chang2002anomaly}, anomaly pixels are considered in the right-tail of the distribution of the MD scores and thus automatic thresholding method has been proposed. The distribution is empirically obtained by normalizing the histogram of the MD scores. Then, the probability of detection is given by $\gamma = P(\{\mathrm{MD(\mathbf{f})}>\alpha \})$ where $\alpha$ is a threshold value which controls the anomaly assignment and $\gamma$ is the confidence coefficient for background. By setting $\gamma$ to a specific value, the threshold $\alpha$ can be calculated. If the MD score is greater than $\alpha$, the test pixel is labeled as an anomaly, otherwise it is labeled as a background pixel. After this step, we obtain a binary detection map in which the anomalies are labeled as 1 and the background is labeled as 0. We use the pixels labeled as the background to constitute our training samples for the AEAN models. In the 1D-AEAN model, we use the $L$-dimensional spectral vectors, $\mathbf{s}_{i}^{(1)}$, in the background class to construct training set $\mathcal{S}^{(1)}=\{\mathbf{s}_{1}^{(1)},\ldots,\mathbf{s}_{N_1}^{(1)}\}$. In the 2D-AEAN model, we extract $m \times m$ small single-band image blocks, $\mathbf{s}_{i}^{(2)}\in \mathbb{R}^{m \times m}$, from the background region by moving a sliding window from left to right and from top to bottom with 8-pixel step size and use them to construct the training set $\mathcal{S}^{(2)} = \{\mathbf{s}_{1}^{(2)},\ldots,\mathbf{s}_{N_2}^{(2)}\}$. In the 3D-AEAN model, we extract $m \times m \times L$ image cubes, $\mathbf{s}_{i}^{(3)}\in \mathbb{R}^{m \times m \times L}$, from the background region to construct the training set $\mathcal{S}^{(3)} = \{\mathbf{s}_{1}^{(3)},\ldots,\mathbf{s}_{N_3}^{(3)}\}$.

\subsection{Reconstruction Step: Model Training}

Adversarial learning-based model called generative adversarial networks (GANs) is first proposed in \cite{goodfellow2014generative}. The GAN model consists of two sub-networks called generator and discriminator. The generator sub-network is employed to synthesize fake input data by mapping the noisy vector $\mathbf{z} \sim f_{\mathbf{Z}}(\mathbf{z})$ in hidden space to $\mathbf{s} \sim f_{\mathbf{S}}(\mathbf{s})$ in the original data space. The discriminator sub-network is used to distinguish between the real and the fake data. We replace the generator sub-network of GAN by an autoencoder to obtain our proposed AEAN model. The most similar models to our proposed one can be found in \cite{sabokrou2018adversarially} and \cite{akcay2018ganomaly}. The autoencoder and the discriminator sub-networks of the AEAN model is demonstrated in Fig. \ref{ganmodel}. The encoder and decoder layers of the autoencoder sub-network are designed by using convolutional and deconvolutional layers, respectively.
\begin{table}
\caption{2D-AEAN architectures. Parameter notation: kernel width $\times$ kernel height $\times$ number of input channels $\times$ number of output channels.}
\renewcommand{\arraystretch}{1.2}
\centering
\begin{tabular}{c|cccc}
\hline
Nets& Layer &  Parameters&  BN & Activation \\
\hline
\hline
\multirow{3}{5em}{Encoder} & Conv  & 9x9x1x64& YES& LReLU  \\
             & Conv & 5x5x64x128& YES& LReLU  \\
             & Conv  & 3x3x128x256& YES& LReLU  \\

\hline            
\multirow{3}{5em}{Decoder}  & 
Deconv& 3x3x256x128& YES& LReLU  \\
             & Deconv  & 5x5x128x64& YES& LReLU  \\
             & Deconv & 9x9x64x1& NO& Tanh \\
\hline
\multirow{4}{5em}{Discriminator}  & Conv& 9x9x1x64& YES& LReLU   \\
             & Conv  & 5x5x64x128& YES& LReLU  \\
             & Conv & 3x3x128x256& YES& LReLU  \\
             & Linear & 256x1& NO& Sigmoid  \\
 \hline
\end{tabular}
\label{2darc}
\end{table}
The network architectures of the 2D-AEAN model is given in Table \ref{2darc}. Unlike the 2D-AEAN model, in the 1D-AEAN model, we set the height of kernels to 1. In the 3D-AEAN model, the number of input channels for the first layer of the encoder and the discriminator, and the number of the output channels for the last layer of the decoder are set to be $L$ instead of $1$. The input data are $L \times 1$ spectral vectors, $m \times m$ image blocks, and $m \times m \times L$ small image cubes for the 1D-AEAN, 2D-AEAN and 3D-AEAN models, respectively.

Since we use three different network architectures, we denote the autoencoder and discriminator functions of the networks as $A_{d}$ and $D_{d}$, respectively, where $d\in\{1,2,3\}$ indicates the dimension of the input data. During the training phase, the autoencoder and the discriminator sub-networks are jointly optimized. To train the model, we use the linear combination of the adversarial and reconstruction losses. In the adversarial loss, we feed the autoencoder with input sample $\mathbf{s}^{(d)}$. The output layer of the AEAN model is a logistic sigmoid. Therefore, the natural loss function ${L}_{Adv}$ between the input and the output is cross-entropy given by
\begin{equation}
\begin{split}
\mathcal{L}_{Adv}(A_{d},D_{d})=\mathrm{E}_{\mathbf{s}^{(d)}} [\log(D_{d}(\mathbf{s}^{(d)}))] + \\
                        \mathrm{E}_{\mathbf{s}^{(d)}} [\log(1-D_{d}(A_{d}(\mathbf{s}^{(d)})))].
\end{split}
 \label{eq:gan}
\end{equation}
We also include a reconstruction loss function $\mathcal{L}_\mathcal{R}$ to the objective function in (\ref{eq:gan}). The loss function $\mathcal{L}_\mathcal{R}$ is a regularization constraint for autoencoder which provides that the error between the output and input of autoencoder is minimum. We choose $\ell_1$-norm loss function because it is more robust to outliers compared to conventional $\ell_2$-norm loss function. The reconstructed sample data is calculated by $\mathbf{\hat{\mathbf{s}}}^{(d)} = A_{d}(\mathbf{s}^{(d)})$ and, then, the reconstruction loss function is found by $ \mathcal{L}_\mathcal{R} = \mathrm{E}_{\mathbf{s}^{(d)}} \left[ \|  \mathbf{s}^{(d)}-\mathbf{\hat{\mathbf{s}}}^{(d)}\|_1 \right]$.
Finally, the learned networks $\hat{A}_{d}$ and $\hat{D}_{d}$ are found by optimizing the total loss function as follows:
\begin{equation}
\label{csmp2}
\{\hat{A}_{d},\hat{D}_{d}\} = \mathrm{arg} \min_{A_{d}}\max_{D_{d}}  \mathcal{L}_{Adv}+ \lambda \mathcal{L}_\mathcal{R},
\end{equation}
where the parameter $\lambda$ controls the balance between two loss functions.

After learning the parameters of the AEAN models, we measure pixel-wise reconstruction error between real and synthesized HSIs. In the reconstruction step, we generate synthesized HSI by using the autoencoder sub-networks in the AEAN models. In the 1D-AEAN, we feed $L \times 1$ spectral vectors to the input of the autoencoder. The autoencoder sub-network reconstruct $L \times 1$ spectral vectors at the output. After the reconstruction of all the spectral vectors in the image, we assemble all the reconstructed vectors to obtain the synthesized HSI. In the  2D-AEAN, we divide the image into non-overlapping (distinct) $m \times m$ blocks and reconstruct each individual block with autoencoder. Unlike in the training phase, in the reconstruction phase, the spatial order of the blocks is important. Thus, we place the reconstructed blocks in the corresponding locations in the synthesized HSI. We apply a similar procedure for 3D-AEAN by using $m \times m \times L$ image cubes instead of $m \times m$ blocks. We define $i$th spectral vector in synthesized HSI as $\hat{\mathbf{f}}_i$. After the reconstruction step, we measure pixel-wise reconstruction error between real and synthesized HSIs using squared Euclidean distance. The reconstruction error is calculated for the $i$th pixel as $r_i = \|\mathbf{f}_i-\hat{\mathbf{f}}_i\|_{2}^2$. The REM value for each pixel is obtained after the reconstruction error calculation step. We form $M_1 \times M_2$ sized REM $R\in \mathbb{R}^{M_1 \times M_2}$ using $r_i$'s. In order to obtain a smooth REM, we apply a morphological filter to the REM as a post-processing. The filter is designed with a closing operator that removes the small holes in the map.

\subsection{Detection Step: WRX-based Detector}

In order to perform a pixel-wise anomaly detection, we use a WRX-based detector assuming that the background pixels follow a multivariate Gaussian distribution. The WRX detector was proposed in \cite{guo2014weighted} to reduce the contribution of the anomaly pixels to the background estimation. Unlike the original WRX, \cite{guo2014weighted}, in this paper, we calculate the weights using the REM values rather than using spectral vectors. We expect that the background pixels are reconstructed with a lower error than the anomaly pixels in the test phase. Therefore, high weights will be assigned to those pixels. 

After obtaining smooth REM, $\tilde{r}_i$, the weight of the $i$th pixel is calculated as $w_{i}=\frac{1}{\tilde{r}_i}$. We normalize the weights as $\bar{w}_{i}=\frac{w_i}{\sum_{i=1}^{N} w_i}$. After the normalization step, all the weights sum to 1 and the weighted mean vector $\hat{\mathbf{m}}$ and the weighted covariance matrix $\hat{\mathbf{C}}$ introduced in \cite{guo2014weighted} can be found as follows: $\hat{\mathbf{m}} =\sum_{i=1}^{N}\bar{w}_{i}\mathbf{f}_i$ and $\hat{\mathbf{C}} =\sum_{i=1}^{N}\bar{w}_{i} (\mathbf{f}_i-\hat{\mathbf{m}} )(\mathbf{f}_i-\hat{\mathbf{m}} )^{T}$. After the parameter estimation step, the anomaly score function is calculated by $\mathrm{AD}(\mathbf{f}_i)=(\mathbf{f}_i-\hat{\mathbf{m}} )^{T}\hat{\mathbf{C}}^{-1}(\mathbf{f}_i-\hat{\mathbf{m}} )$.

\section{Experimental Results}
\label{SecExp}  
In this section, we evaluate the detection performances of our proposed method and compare it with those of the RX \cite{RX90}, WRX \cite{guo2014weighted}, SVDD \cite{banerjee2006support}, CRD \cite{li2014collaborative}, AW-DBN \cite{ma2018unsupervised} detectors and DAEAD \cite{taghipour2019unsupervised} method. The detection performances are measured with the area under the receiver operating characteristic (ROC) curve (AUC) metric. 

\begin{figure}[t]
  	\centering{
 	\subfloat[ABU Airport1]{\includegraphics[width=1.40cm]{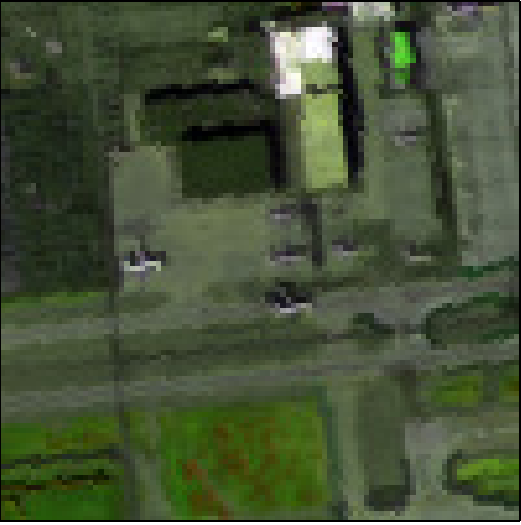}
 		\label{abuair1_im} \hspace{-0.2cm}
 		\includegraphics[width=1.40cm]{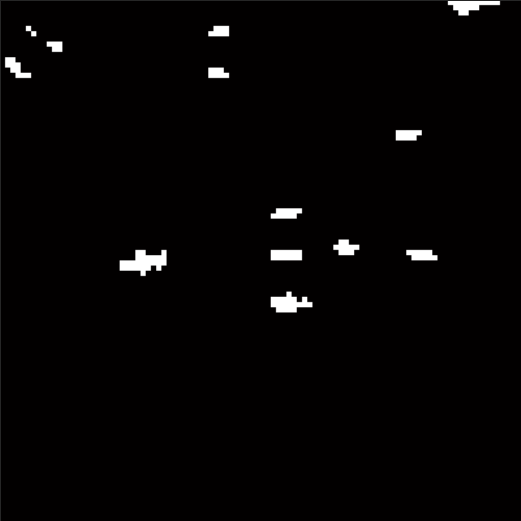}} \hspace{-0.05cm}
 	\hspace{-0.1cm}
 	 \subfloat[ABU Urban3]{\includegraphics[width=1.40cm]{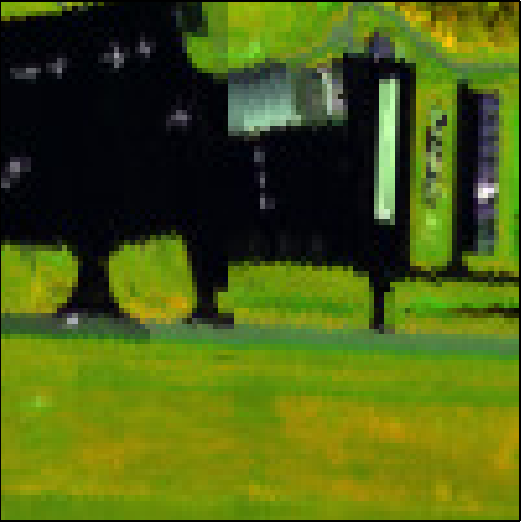}
 		\label{abuurban2_im} \hspace{-0.2cm}
 		\includegraphics[width=1.40cm]{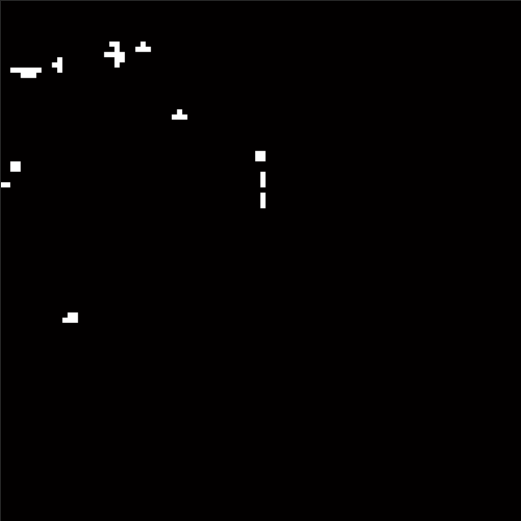}} 
 		\hspace{-0.05cm}
 	 	 \hspace*{-0.1cm}
 	 \subfloat[San Diego]{\includegraphics[width=1.40cm]{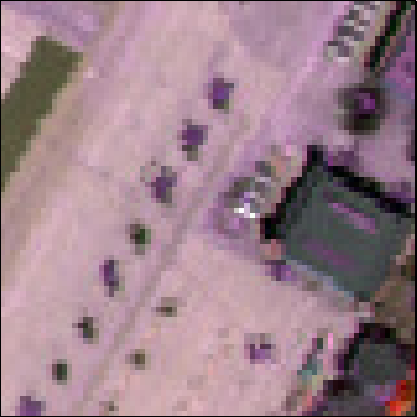}
 		\label{sandiego_im} \hspace{-0.2cm}
 		\includegraphics[width=1.40cm]{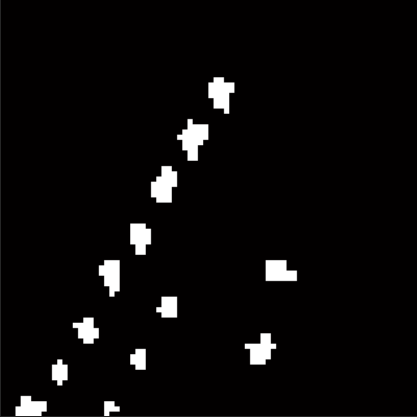}} 
     		}
  	\caption{Three-band color composites and reference maps.}
  	\label{ImageGT}
  \end{figure}

\begin{figure}
	\centering{
		\includegraphics[width=8.0cm]{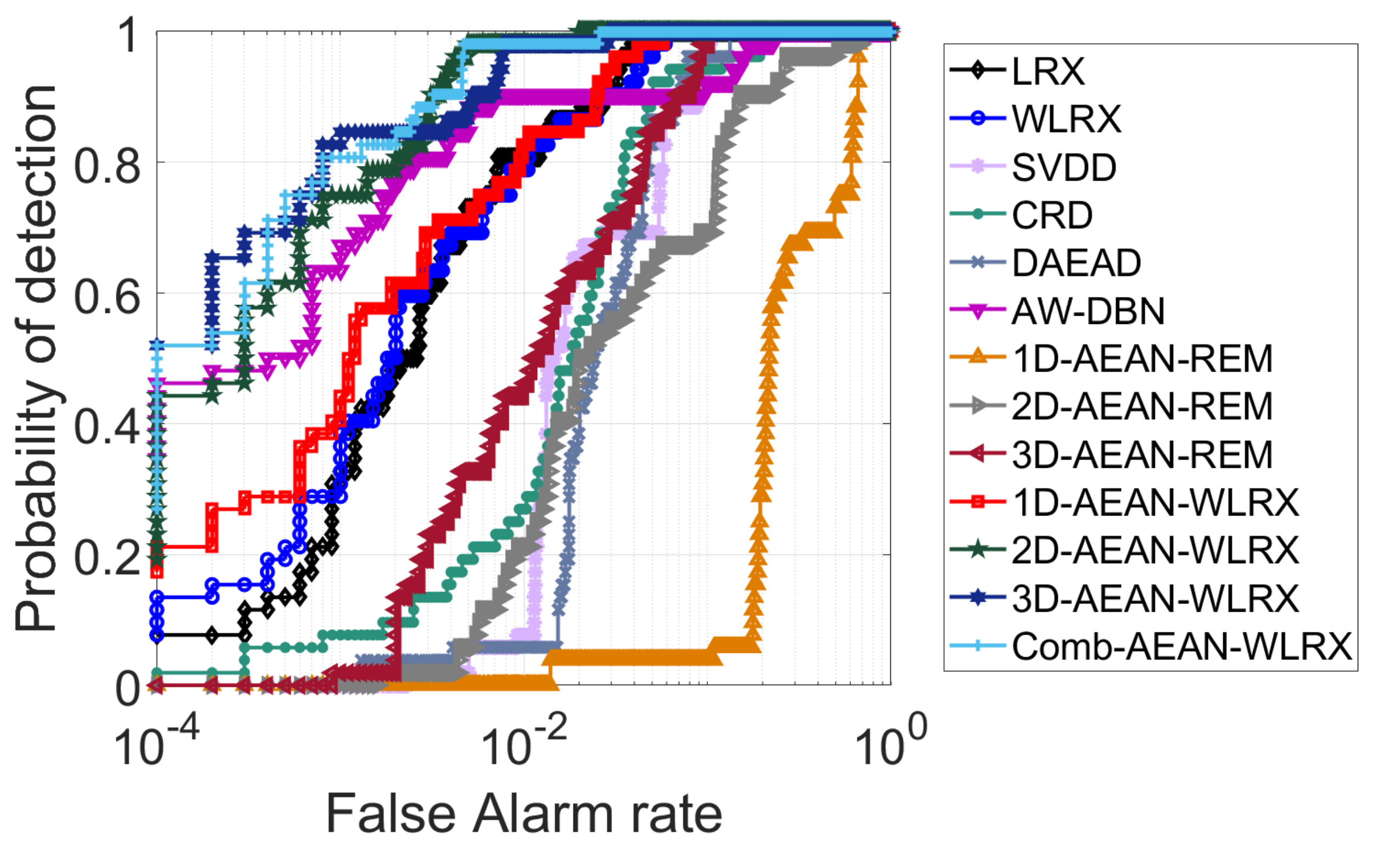}
	}%
	\caption{ROC curve for the ABU Urban3.}
	\label{ROCcurve}
\end{figure}

We use real HSIs for performance evaluation. The authors of \cite{kang2017hyperspectral} manually created the ABU (Airport Beach Urban) dataset by using HSIs captured by the Airborne Visible/Infrared Imaging Spectrometer (AVIRIS) sensor, except the ABU Beach4 image, which is captured by the Reflective Optics System Imaging Spectrometer (ROSIS)-03 sensor. The spatial sizes of the images are within the range between $100\times 100$ and $150\times 150$. The numbers of spectral bands are in the range between $102$ and $205$. In the Airport scene, the airplanes are defined as anomalies, in the urban scene, the small buildings are defined as anomalies and in the beach scene, the boat on the sea or the vehicles on the bridge are considered as anomalies. We also use the San Diego image captured by the AVIRIS sensor. The original image size is $400\times 400\times 224$ and the dataset can be downloaded from \cite{WinNT}. After the noisy bands are removed, only 189 spectral bands are used. We cut $80\times 80\times 189$ image from the region with airplanes. The airplanes are assumed to be as anomalies. Three-band color composite and reference maps of some images are shown in Fig. \ref{ImageGT}. In the data preparation step, we empirically set the confidence coefficient $\gamma$ to a value between $0.97$ and $0.9999$. The detection performance is not changed significantly by the values of $\gamma$. The maximum change in AUC, in some images, is about $\pm 0.01$. We normalize the intensity of each HSI to range $[-1,1]$ and set the balance term $\lambda$ in (\ref{csmp2}) to $10$. We also investigate different $\lambda$ values and observe that the computation time during training increases when we set $\lambda$ to a small value, but the detection result is not significantly changed. The block size is set to 16 for the 2D-AEAN and 3D-AEAN models. In the local version, we use a dual window approach and choose the inner and outer window sizes according to the AUC metric. During the experiments, we generally observe that the best AUC result is obtained when we set the inner window size to 1. Similarly, the parameter $\sigma$ of SVDD, the window size of the local RX (LRX), the weighted local RX (WLRX), and the CRD are selected based on the AUC metric. In local methods, we also regularize the covariance matrix estimation using the proposed method in \cite{nasrabadi2008regularization}. The value of the regularization coefficient is selected according to the AUC metric for each local method.
Since the network weights in DAEAD, DBN and AEAN models depend on initial parameters, we run the learning algorithms 10 times and calculate average detection results. In local detection, we call our proposed methods as 1D-AEAN-WLRX, 2D-AEAN-WLRX, and 3D-AEAN-WLRX. We also analyze the detection performances in case that the REM values are directly used as anomaly scores. These methods are denoted as 1D-AEAN-REM, 2D-AEAN-REM, and 3D-AEAN-REM. We also obtain Comb-AEAN-WLRX method by combining the anomaly scores of the 1D-AEAN-WLRX, 2D-AEAN-WLRX and 3D-AEAN-WLRX. The scores of 1D-, 2D- and 3D-AEAN-WLRX are weighted by 0.01, 0.5 and 0.49, respectively, and averaged. The weights are chosen to be proportional to the individual detection performances of the methods. Experimental results based on the AUC metric are presented in Table \ref{tablecomp:all}. The detection performances of simple REM-based detectors are lower than the proposed WLRX-based detectors. As seen from the table, our proposed AEAN-based WLRX detectors outperform the classical LRX and WLRX detectors and deep-learning-based AW-DBN and DAEAD methods. The detection performance of the DAEAD method is also low compared to our proposed AEAN models because only the spectral information is considered during reconstruction and detection steps. We observe in Table \ref{tablecomp:all} that the use of only the spectral information in the AEAN model is not sufficient to model the background and, as a result, to discriminate the anomaly pixels as well. As seen from Table \ref{tablecomp:all}, the 2D-AEAN-WLRX and 3D-AEAN-WLRX detectors outperform the other compared methods. It is seen from the ROC curve for the ABU Urban3 in Fig. \ref{ROCcurve} that 2D-AEAN-WLRX, 3D-AEAN-WLRX and Comb-AEAN-WLRX detectors provide higher detection probability with low false alarm rate. 

\begin{figure}[!ht]
 	\centering{
		\subfloat[\scriptsize{Image}]{\includegraphics[width=1.7cm]{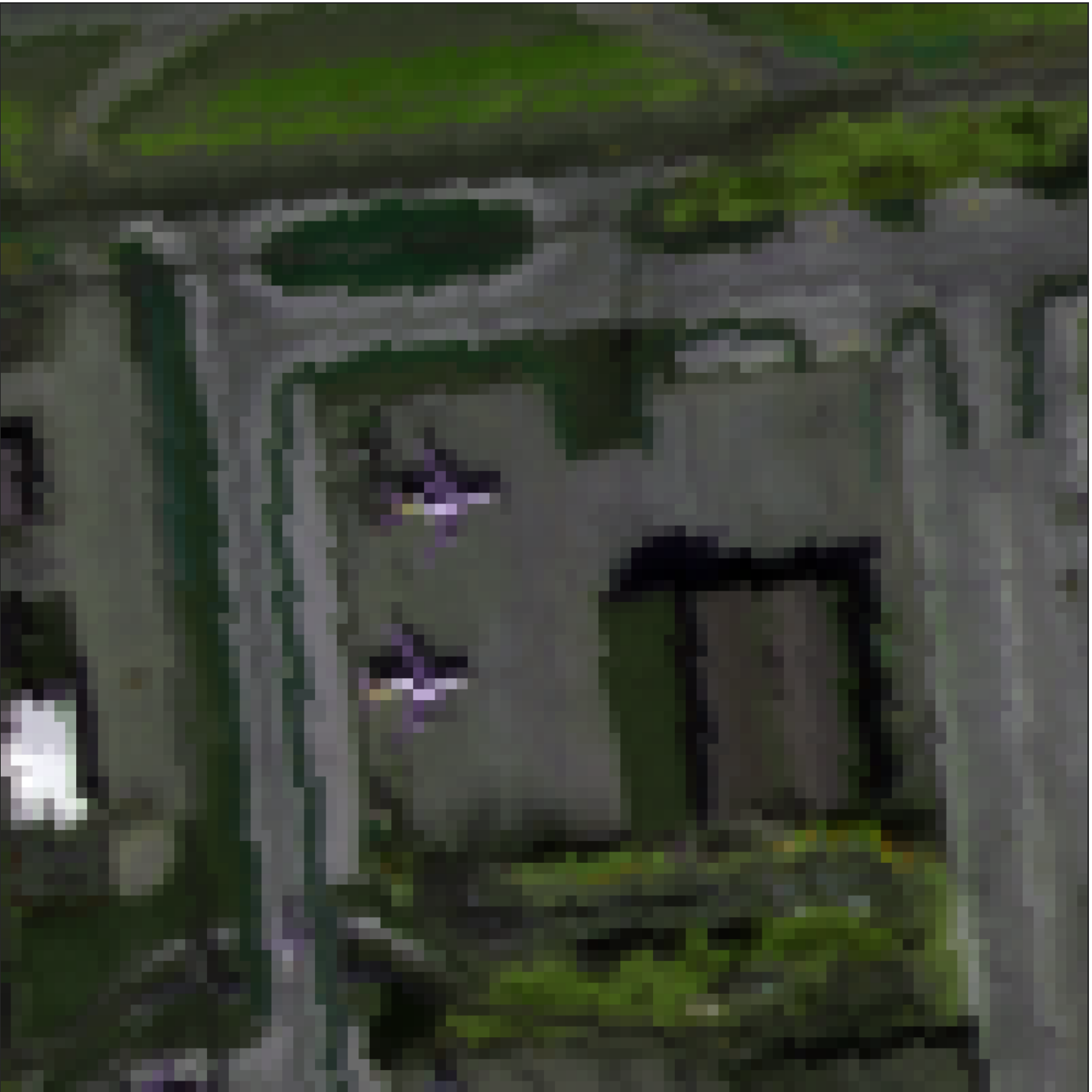}
 			\label{abu_air2}} 
 		\subfloat[\scriptsize{Reference}]{\includegraphics[width=1.7cm]{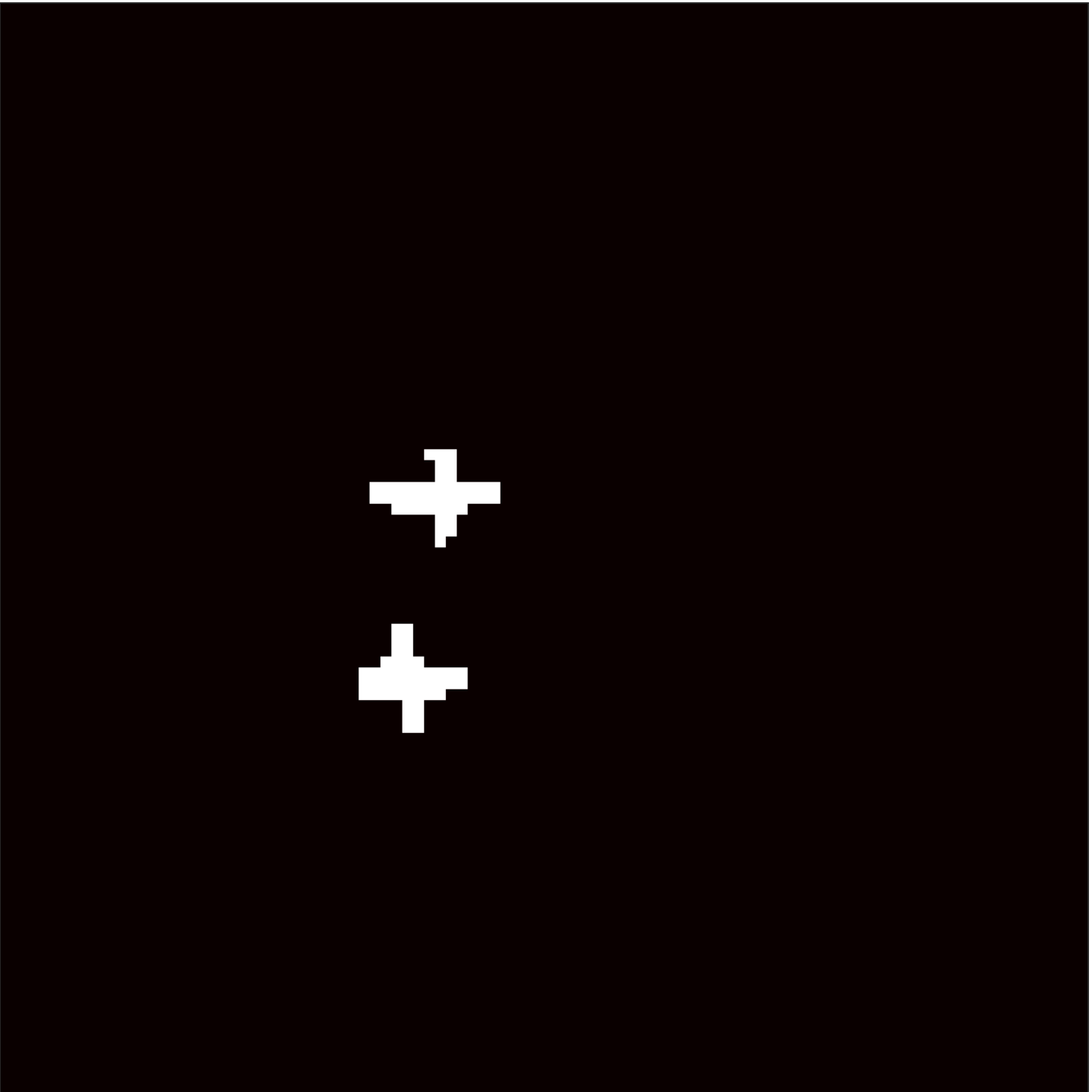}
 			\label{abu_air_gt}} 
 		 \subfloat[\scriptsize{LRX}]{\includegraphics[width=1.7cm]{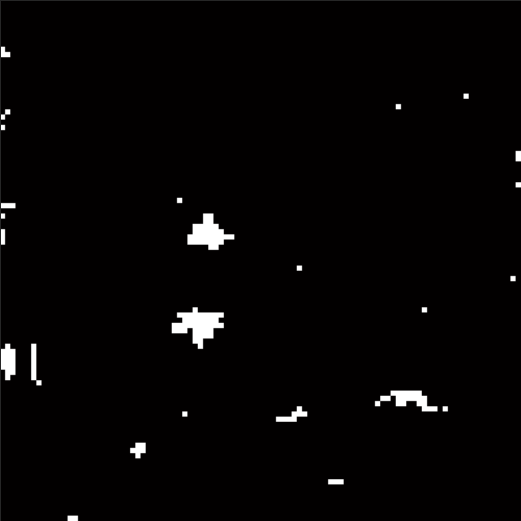}
 			\label{abu_air2_lrx}} 
 		\subfloat[\scriptsize{WLRX}]{\includegraphics[width=1.7cm]{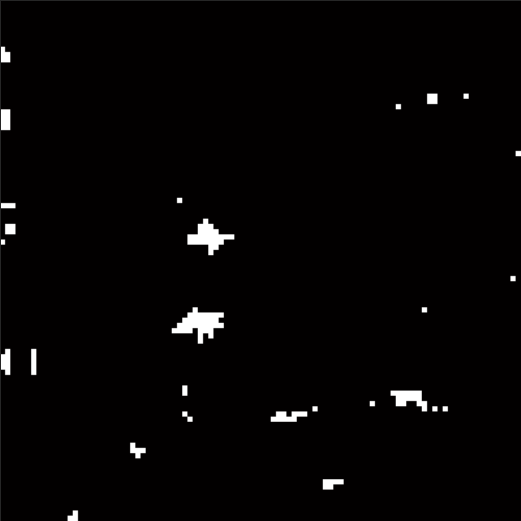}
 			\label{abu_air2_wlrx}} 
 		\subfloat[\scriptsize{CRD}]{\includegraphics[width=1.7cm]{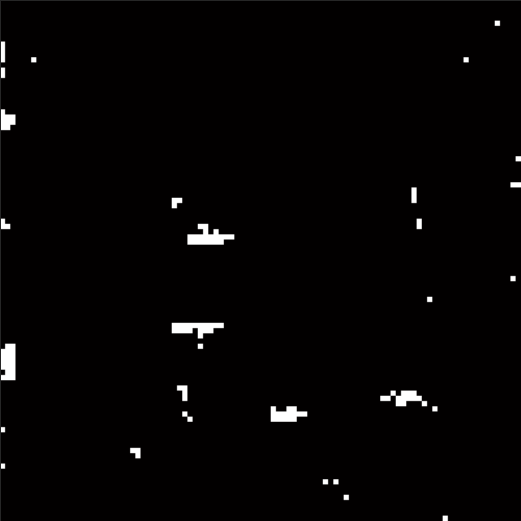}
 			\label{abu_air2_crd}} 
 			\vspace{-0.2cm}
 					\\
 		\subfloat[\scriptsize{SVDD}]{\includegraphics[width=1.7cm]{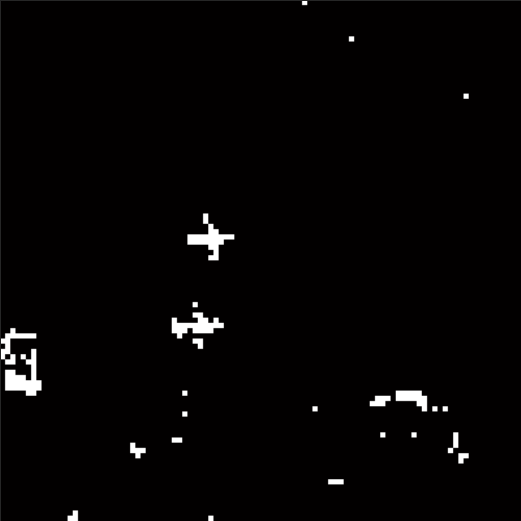}
          \label{abu_air2_svdd}} 
 	 	\subfloat[\scriptsize{DAEAD}]{\includegraphics[width=1.7cm]{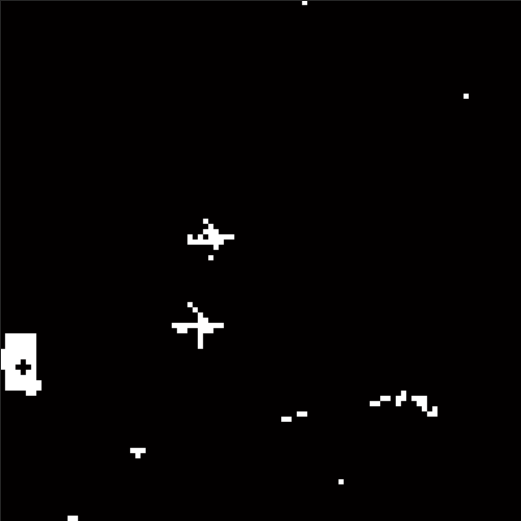}
 		\label{abu_air2_daead}} 
 		 \subfloat[\scriptsize{AW-DBN}]{\includegraphics[width=1.7cm]{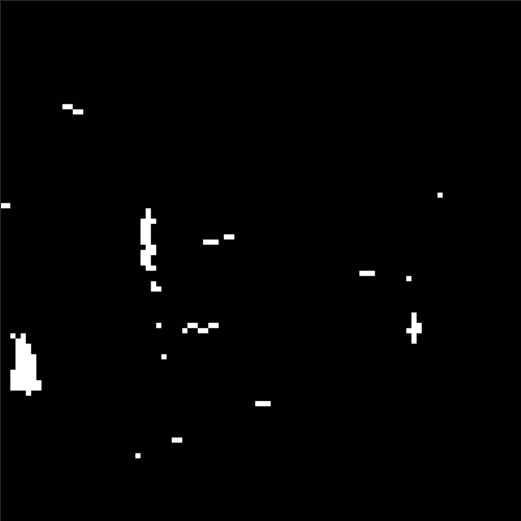}
 		\label{abu_air2_awdbn}} 
 		\subfloat[\scriptsize{1D-AEAN-REM}]{\includegraphics[width=1.7cm]{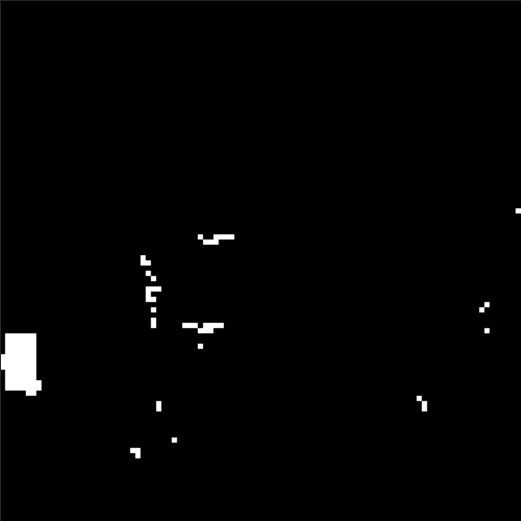}
 		\label{abu_air2_1daean_rem}} 
 		\subfloat[\scriptsize{2D-AEAN-REM}]{\includegraphics[width=1.7cm]{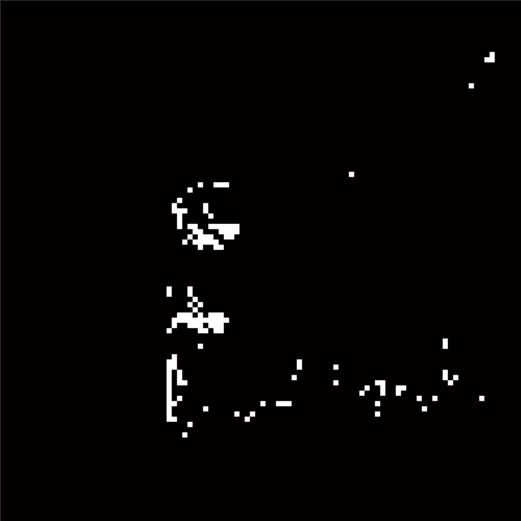}
 		\label{abu_air2_2daean_rem}} 
 			\vspace{-0.2cm}
 		\\
 		\subfloat[\scriptsize{3D-AEAN-REM}]{\includegraphics[width=1.7cm]{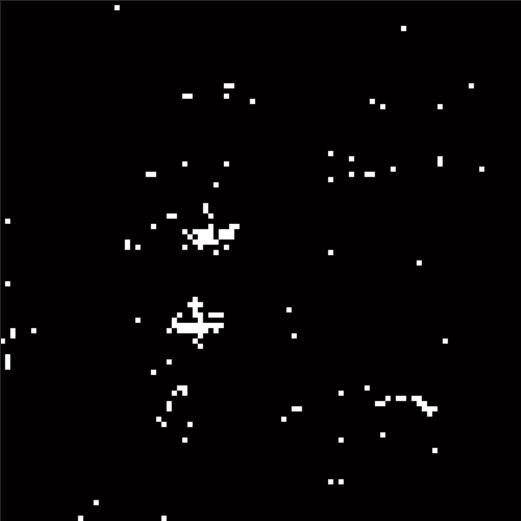}
 		\label{abu_air2_3daean_rem}} 
 	 	\subfloat[\scriptsize{1D-AEAN-WLRX}]{\includegraphics[width=1.7cm]{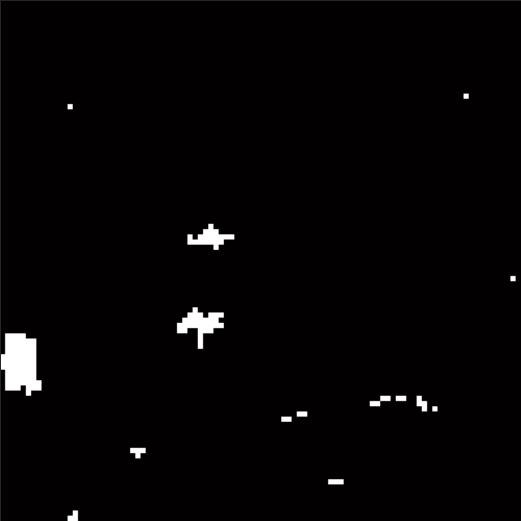}
 		\label{abu_air2_1daean_wlrx}} 
 		\subfloat[\scriptsize{2D-AEAN-WLRX}]{\includegraphics[width=1.7cm]{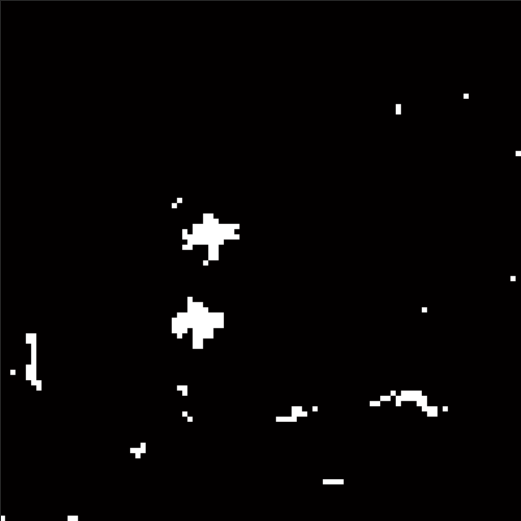}
 		\label{abu_air2_2daean_wlrx}} 
 		\subfloat[\scriptsize{3D-AEAN-WLRX}]{\includegraphics[width=1.7cm]{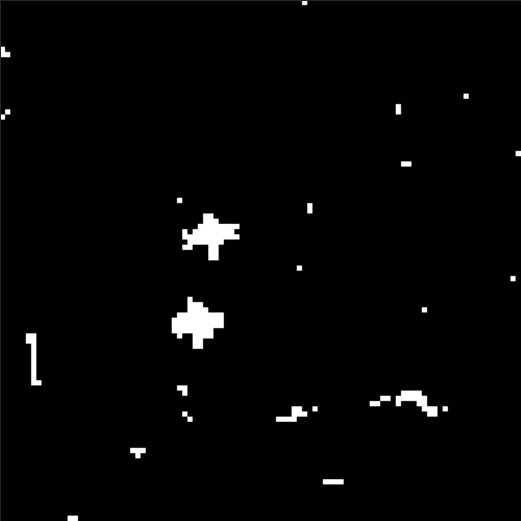}
 		\label{abu_air2_3daean_wlrx}} 
 		\subfloat[\scriptsize{Comb-AEAN-WLRX}]{\includegraphics[width=1.7cm]{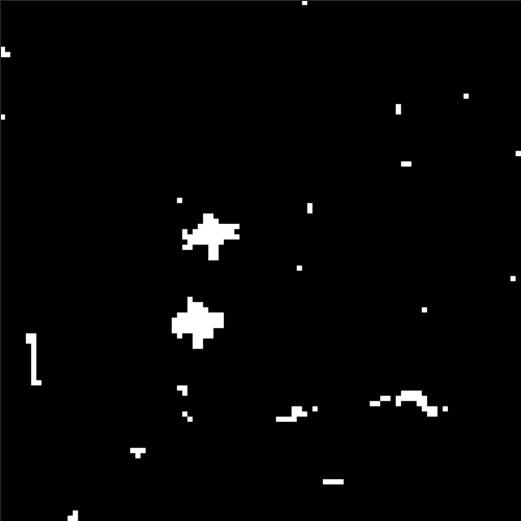}
 		\label{abu_air2_combaean_wlrx}} 
 }
\caption{Detection maps for ABU Airport2 at false alarm rate 0.01.}
\label{fig:dm}
\end{figure}

According to our experiments, the use of the morphological filter on REM increases the detection performance of 1D-, 2D-, and 3D-AEAN-WLRX AD by 0.0146, 0.0081 and 0.0055 on average, respectively. Fig. \ref{fig:dm} shows the detection maps for the ABU Airport2 at false alarm rate 0.01. As seen from Fig. \ref{fig:dm}, the detection maps obtained by local detectors are better than the global detectors. The 2D-AEAN-WLRX and 3D-AEAN-WLRX detection maps are closer to the reference map and the shapes of the anomalies are detected more accurately.

In the last row of Table \ref{tablecomp:all}, we give the average computation times of the methods for the ABU Urban1 image. Since the training times of the deep networks are naturally higher compared to classical methods, we give the computation time of detection. As seen from the table, global and REM-based detectors are faster than local detectors. Since the weights are calculated in the WLRX-based detectors, the LRX is faster than the WLRX-based detectors but there is not significant difference in the computation time between the WLRX-based and local detectors. The detection performance is increased about 0.0007 by the Comb-AEAN-WLRX in only four images, but the computation time is increased 3 times.
\begin{table*}[ht!]
\caption{AUC results and average computation time of the methods in seconds.}
	\begin{center}
			\renewcommand{\arraystretch}{1.1}

		\begin{tabular}{p{1.0cm}p{0.8cm}p{0.8cm}p{0.8cm}p{0.8cm}p{0.8cm}p{0.8cm}p{0.8cm}p{0.8cm}p{0.8cm}p{0.8cm}p{0.8cm}p{0.8cm}p{0.8cm}}
			\hline
			Images  & LRX  & WLRX  &  CRD &  SVDD  & DAEAD & AW-DBN & 1D-AEAN-REM & 2D-AEAN-REM & 3D-AEAN-REM & 1D-AEAN-WLRX & 2D-AEAN-WLRX & 3D-AEAN-WLRX & Comb-AEAN-WLRX \\
			\hline
			\hline
			Airport1  & 0.9703 & 0.9692 & 0.9663 & 0.9454 & 0.8994 & 0.9164 & 0.6945 & 0.8763 & 0.9186 & 0.9614 & 0.9731 & \textbf{0.9747} & \underline{0.9763} \\
			Airport2  & 0.9895 & 0.9878 & 0.9685 & 0.9759 & 0.9166 & 0.8330  & 0.7251 & 0.9731 & 0.9681 & 0.9805 & \textbf{0.9962} & 0.9955 & 0.9959 \\
			Airport3  & 0.9608 & 0.9566 & 0.9422 & 0.9645 & 0.9322 & 0.9066 & 0.8390 & 0.8506 & 0.8983 & 0.9651 & 0.9825 & \textbf{0.9829} & \underline{0.9832} \\
			Airport4  & 0.9942 & 0.9939 & 0.9862 & 0.9891 & 0.9493 & 0.9893 & 0.8682 & 0.9681 & 0.9843 & 0.9928 & \textbf{0.9950} & 0.9935 & 0.9943 \\
			Urban1    & 0.9971 & 0.9971 & \textbf{0.9978} & 0.9954 & 0.9868 & 0.9934 & 0.9865 & 0.9824 & 0.9766 & \textbf{0.9978} & 0.9974 & 0.9973 & \underline{0.9979} \\
			Urban2    & 0.9982 & 0.9959 & 0.9842 & \textbf{0.9993} & 0.9907 & 0.9991 & 0.9991 & 0.9916 & 0.9977 & 0.9989 & 0.9985 & 0.9988 & 0.9987 \\
			Urban3    & 0.9929 & 0.9920 & 0.9713 & 0.9689 & 0.9524 & 0.9832 & 0.7780 & 0.9190 & 0.9712 & 0.9925 & \textbf{0.9984} & 0.9983 & \underline{0.9986} \\
			Urban4    & 0.9942 & 0.9947 & 0.9816 & 0.9948 & 0.9551 & 0.9944 & 0.9873 & 0.9763 & 0.9751 & 0.9956 & \textbf{0.9966} & 0.9963 & 0.9965 \\
			Urban5    & 0.9644 & 0.9647 & 0.9521 & 0.9732 & 0.9592 & 0.9718 & 0.9534 & 0.9552 & 0.9448 & 0.9839 & \textbf{0.9876} & 0.9855 & 0.9872  \\
			Beach1    & \textbf{0.9986} & 0.9985 & 0.9970 & 0.9848 & 0.9747 & 0.9974 & 0.9527 & 0.9838 & 0.9894 & 0.9985 & 0.9982 & 0.9981 & 0.9981 \\
			Beach2    & 0.9696 & 0.9691 & 0.9661 & 0.9282 & 0.8936 & 0.9746 & 0.8686 & 0.9449 & 0.9317 & 0.9807 & \textbf{0.9899} & 0.9871 & 0.9889 \\
			Beach3    & \textbf{1.0000} & \textbf{1.0000} & 0.9998 & 0.9989 & 0.9921 & 0.9999 & 0.9501 & 0.9992 & 0.9933 & \textbf{1.0000} & \textbf{1.0000} & \textbf{1.0000} & 1.0000\\
			Beach4    & 0.9892 & 0.9787 & 0.9553 & 0.9803 & 0.9583 & 0.9817 & 0.8816 & 0.9932 & \textbf{0.9936} & 0.9788 & 0.9921 & 0.9911 & 0.9936 \\
			S. Diego     & 0.9748 & 0.9712 & 0.9444 & 0.9029 & 0.7954 & 0.9732 & 0.7572 & 0.7963  & 0.8280 & 0.9726 & 0.9822 & \textbf{0.9897} & 0.9872 \\
		\hline \hline
		Time  & 36.819 & 38.771 & 45.004 & 39.717 & 0.025 & 0.490 & 0.024 & 0.026 & 0.025 & 39.173 & 39.047 & 38.888 & 126.329\\
        \hline
		\end{tabular}
	\end{center}
	\label{tablecomp:all}
\end{table*}

\captionsetup[subfigure]{labelformat=empty}

\section{Conclusion}
\label{SecCon}

In this paper, we propose a completely unsupervised pixel-wise anomaly detection method for hyperspectral images based on AEAN models. We propose three AEAN models each of which work on spectral, spatial, and spectral-spatial domains. As seen from the experimental results, the inclusion of spatial information improves the detection performance. In the detection part, we propose to use the WRX-based anomaly detector to obtain a pixel-wise detection. The WRX-based detection performs better than the REM-based detection. The 2D-AEAN-WLRX and 3D-AEAN-WLRX detectors yield higher detection performance than the other detectors in the benchmark. While in 2D-AEAN, spectral and spatial data are independently processed, in 3D-AEAN, they are jointly processed. The proposed method can be adapted for sub-pixel hyperspectral anomaly detection. The depth of the network can be increased using the residual network structure and more hyperspectral data for training. 

\ifCLASSOPTIONcaptionsoff
  \newpage
\fi



%

\bibliographystyle{IEEEtran}


%








\end{document}